# Study of Encryption and Decryption of Wave File in Image Formats

**Rahul R Upadhyay**
Department of Mechanical Engineering, BBD National Institute of Technology and Management, Lucknow
Email: rahulrupadhyay91@gmail.com

-------------------------------------------------------------------------**ABSTRACT**--------------------------------------------------------------------
This paper describes a novel method of encrypting wave files in popular image formats like JPEG, TIF and PNG along with retrieving them from these image files. MATLAB software is used to perform matrix manipulation to encrypt and decrypt sound files into and from image files. This method is not only a stenographic means but also a data compression technique.

-------------------------------------------------------------------------------------------------------------------------------------------------------------

## 1. INTRODUCTION

This present work puts forth a novel method to encrypt '.wav', which are basically sound files in image formats such as PNG, TIF and JPEG. The sound file is fetched and the values corresponding to the sample range is put in a column matrix which is then put in a two dimensional matrix having "double" as data-type. Using 'imwrite' function of MATLAB, this matrix defined in the class 'double' is put in a graphics file or image file having dynamic range from 0 to 1 [18].

After encryption the data is retrieved from the image file and compared with the original wave file to show the variation in encrypted and decrypted data. This methodology can not only be used as stenographic means but possibly as a technique for data compression. The illustrated method for data encryption of sensitive user information can be used as a viable method to further secure cloud computing transactions [11] [24] [27].

### 1.1 WAVE Bitstream Format

Waveform Audio File Format (WAVE) is an application of RIFF or Resource Interchange File Format which stores audio bit streams in "chunks". WAVE encodes the sound in LPCM format i.e. Linear Pulse Code Modulation [1] [22].

Sound is basically a pressure wave or mechanical energy having pressure variance in an elastic medium. The variance propagates as compression and rarefaction wherein compression occurs when pressure is higher than the ambient pressure and rarefaction occurs when the pressure of the propagating wave is less than the ambient pressure.

Exactly in the same manner a WAVE file just represents the sampled sound waves which happen to be above or below the equilibrium or ambient air pressure.

In this paper we will be using a "drums.wav" wave file to show the proposed algorithm of encrypting the sound file in various image formats [17] [26]. As already mentioned a wave file consists of positive and negative values over its entire range of samples. Here for the sake of simplicity we will be using only the samples having positive values. The sound file will be read and handled in MATLAB.

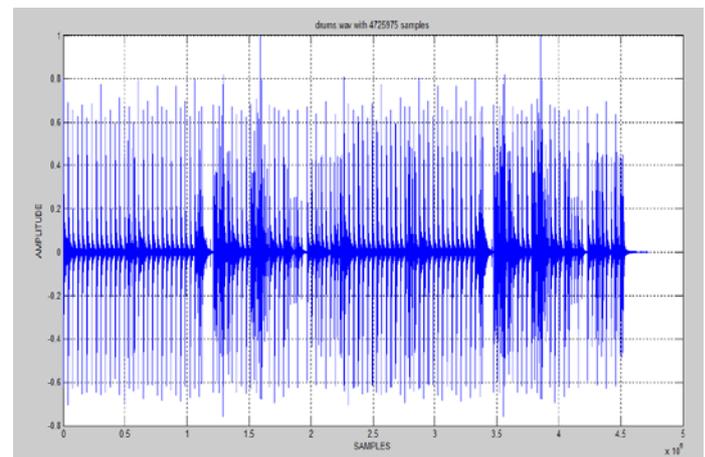

Figure 1 Graphical representation of "drums.wav" wave file in its entire sampling range.

### 1.1 Image Formats and Image Compression

Digital image formats are means of storing digital images in either uncompressed (e.g. TIFF), compressed (e.g. JPEG) and vector formats. On rasterization an image is converted into a grid of pixels. Basically there are two types of image file compression algorithm- lossless and lossy. In lossless compression the entire digital data is preserved during compression thus preserving image quality. In lossy compressions, the digital data preservation takes place by compromising image quality [2]. Here we will be discussing PNG, TIFF and JPEG formats and these are the very formats in which the wave file will be encrypted into, in this paper.

TABLE I
RASTER IMAGE FORMATS

| Sr. no | Image Format | Description in Brief |
|---|---|---|
| 1 | JPEG | JPEG compresses image files to great extent but at the cost of image quality. |
| 2 | TIF | TIF is a lossless compression format which is considered as industry standard. |
| 3 | PNG | PNG files are smaller than TIF(LZW compression) though both are lossless although it is slower to read or write to.[3] |

.

## 2. ENCRYPTION

### 2.1 Obtaining data of wave file in column matrix

We will be using "drums.wav" wave file in this paper whose graphical representation is provided in Fig1. The sampling length of this tone is 4725975 samples. As already mentioned, we will be using only those samples which have positive values. Following is the MATLAB code which fetches the wave file using 'wavread' function [4]. Amplitude values are obtained in the range of 0 and +1.

```
>> % wavread to fetch .wav
>> b=wavread('C:\Users\Rajeeva\Desktop\drums.wav');
>> % C stores all values which are +ve
>> C=b(b>=0);
>> % D stores first 2000000 +ve values
>> D=C(1:2000000);
>> plot(D)
>> grid on
>> xlabel('Samples');
>> ylabel('Amplitude');
```

Figure 2 MATLAB code to fetch wave file and obtain first 2000000 positive samples.

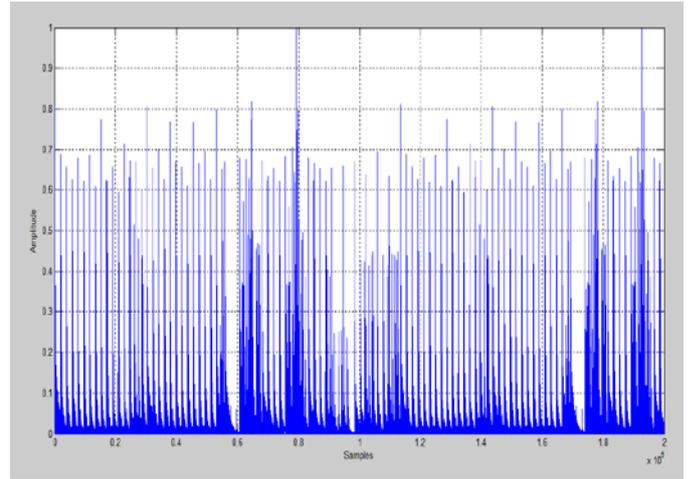

Figure 3 Graphical representation of variable 'D' which holds first 2000000 positive samples. Its value ranges from 0 and +1.

It is to be noted that the variable D is basically a column vector.

### 2.2 Converting column matrix into M x N matrix.

A 'grayscale' [12] image of M by N pixels is represented in MATLAB as an M X N matrix having "double" data type wherein each element of the matrix denotes a pixel within an intensity of 0 and 1. [5]. It is to be noted that the variable D is a column matrix with "double" type and intensity within 0 and 1. So to convert variable D in an image format we have to transform D into a 1000 X 2000 matrix.

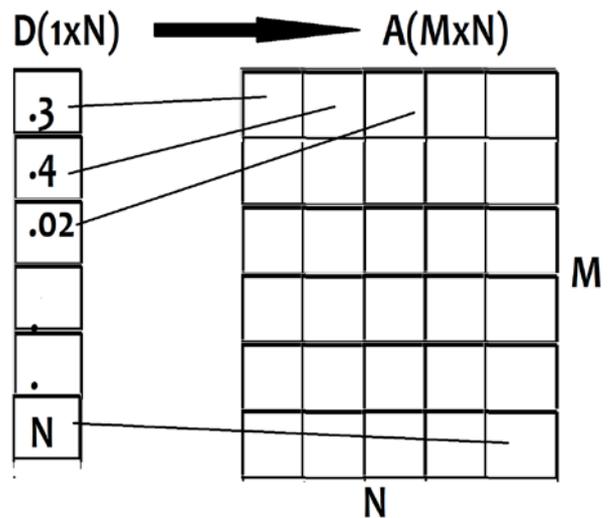

Figure 4 Converting N (1xN) matrix into A (MxN) matrix.

```
Command Window
>> x=1;
>> y=2000;
>> for i=1:1000
% All values in row i are fetched
L=D(x:y);
% Column vector L is transposed
N=L';
% Values are put in Matrix A(2000x1000)
A(i,:)=N;
x=x+2000;
y=y+2000;
end
```

Figure 5 MATLAB code to convert column matrix D into A (MxN) matrix.

### 2.3 Converting matrix into Image File.

We convert matrix A into JPEG [16], PNG and TIF formats using MATLAB function called "imwrite".

**Imwrite (A,'.../filename.xyz'); [6]**

The above syntax stores matrix A in the file path mentioned. We also save column matrix D in a new wave file using "wavwrite" function.

```
>> % converting MxN matrix in PNG, JPEG and TIF
>> imwrite(A, 'C:\Users\Rajeeva\Desktop\Aa.jpg');
>> imwrite(A, 'C:\Users\Rajeeva\Desktop\Aa.png');
>> imwrite(A, 'C:\Users\Rajeeva\Desktop\Aa.tif');
>> wavwrite(D,42100,'C:\Users\Rajeeva\Desktop\Aa.wav');
```

Figure 6 MATLAB code to convert column matrix A in image formats and array D into .wav file.

TABLE II
SIZE OF WAVE FILE AND VARIOUS IMAGE FORMATS IN WHICH THE FILE IS ENCRYPTED INTO

| S.no | NAME | TYPE | SIZE |
|------|------|------|------|
| 1 | Aa | .wav (WAVE) | 3.81 MB |
| 2 | Aa.jpg | JPEG | 293 KB |
| 3 | Aa.tif | TIFF | 1.61 MB |
| 4 | Aa.png | PNG | 734 KB |

It can be clearly seen that JPEG stores the wave file of 3.81 MB into merely 293 KB whereas TIFF stores the same file in 1.61 MB. These images will be studied further in the paper.

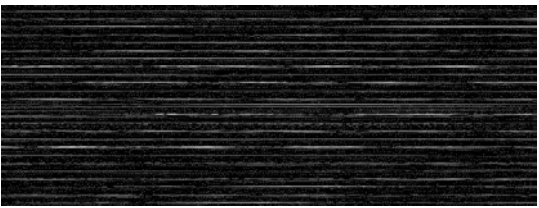

Figure 7 Snapshot of how the wave file appears in JPEG format (this is a cropped image as the actual file is far bigger).

## 3. DECRYPTION OF WAVE FROM IMAGE
(STEGANALYSIS)[13][23] [14]

This method is just the opposite of encryption with minor variations. When the image is created during encryption it is basically an MxN matrix with "double" data type, however on fetching the same image back to MATLAB we get an MxN matrix with datatype "uint8" i.e. unsigned integer of 8 bits [20].

Thus we need to first convert all elements of matrix obtained into double precision [7].

Decryption can be basically understood as a data mining method to fetch, audit and understand the pattern of data stored in the encrypted file[9] [19] .

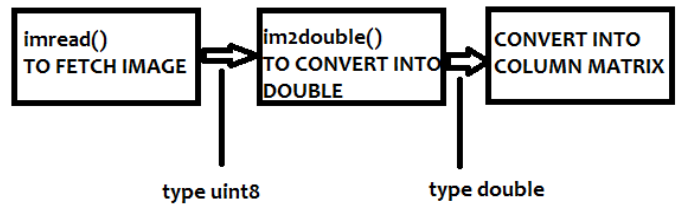

Figure8 Figure showing the necessary conversion of datatype.

### 3.1 Datatype conversion[7]

"**im2double**" function of MATLAB takes an image as input and converts it into another image or matrix having all elements with type double. As already stated, the image files retrieved by MATLAB are in unsigned integer form and before they are put in column matrix representing a WAVE file, it has to be converted into datatype of double precision [21].

```
Command Window
>> %Fetch image file
>> AaJPG=imread('C:\Users\Rajeeva\Desktop\Aa.jpg');
>> AaTIF=imread('C:\Users\Rajeeva\Desktop\Aa.tif');
>> AaPNG=imread('C:\Users\Rajeeva\Desktop\Aa.png');
>> %CONVERT THE ABOVE MATRICES INTO DOUBLE PRECISION
>> AaJPG=im2double(AaJPG);
>> AaTIF=im2double(AaTIF);
>> AaPNG=im2double(AaPNG);
>>
```

Figure 9 MATLAB code to convert datatype of matrices.

### 3.2 ALGORITHM TO CONVERT IMAGE MATRIX INTO COLUMN MATRIX AND HENCE .WAV.[8]

"**wavwrite ( X, FS, '.../filename.wav');**"

The above function is used to save X column vector in the given 'filename' (say DivyaSharma or Shumo) with a desired frequency 'FS'. The column vector X is obtained by converting image matrix of double precision into column matrix. The method is explained in the following programming code of MATLAB.

```
Command Window
>> % defining column matrices
>> JPEG=[;];
>> TIF=[;];
>> PNG=[;];
>> %converting JPG into sound file
>> for i=1:1000
w=AaJPG(i,:);
%transpose of w
m=w';
JPEG=vertcat(JPEG,m);
end
>> i=1;
>> %converting PNG into sound file
for i=1:1000
w=AaPNG(i,:);
%transpose of w
m=w';
PNG=vertcat(PNG,m);
end
>> i=1;
>> %converting TIF into sound file
for i=1:1000
w=AaTIF(i,:);
%transpose of w
m=w';
TIF=vertcat(TIF,m);
end
```

Figure 10 MATLAB Code to convert all image files into Column Matrix/Wave File.

## 4. GRAPHICAL ANALYSIS.

During encryption we used "wavwrite" to store the sample wave file that we used in the paper. We fetch the same wave file and compare it graphically with the sound files that we obtained during decryption.

```
Command Window
>> ORIGINAL=wavread('C:\Users\Rajeeva\Desktop\Aa.wav');
```

Figure 11 Storing the original wave file in variable ORIGINAL.

### 4.1 ORIGINAL WAVE FILE vs. SOUND MATRIX OBTAINED FROM JPEG FILE.

We plot original file with decrypted sound file.

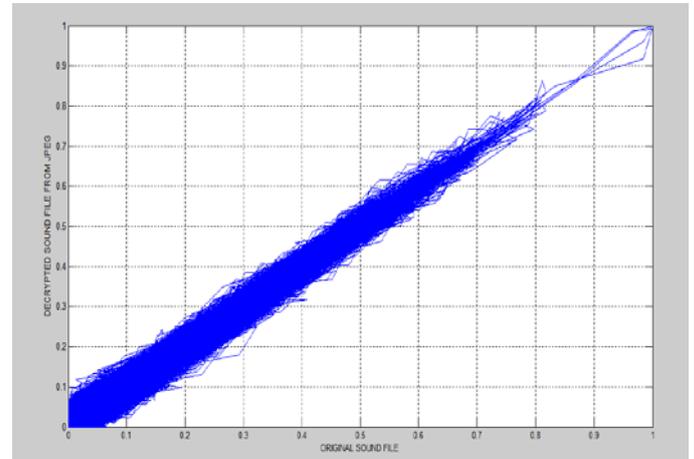

Figure 12 Plot of the original sound file vs. the decrypted sound file.

The graph shows that the image file decrypted using JPEG has decent variation from the actual WAVE file. This variation can be represented by another 1000x2000 matrix called ERROR such that [25]

ERROR = ORIGINAL – JPEG (Equation A)

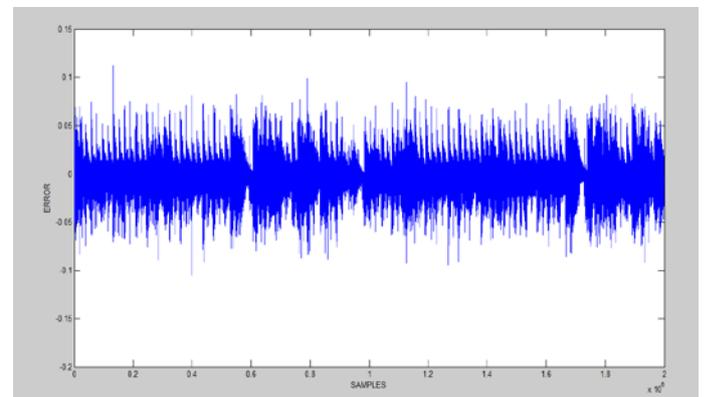

Figure 13 Plot of ERROR vs. SAMPLES.

Error in case of JPEG decrypted wave file varies from -.01 to +.01.

### 4.2 ORIGINAL WAVE FILE Vs. SOUND MATRIX OBTAINED FROM TIF FILE.

Column matrix obtained from TIF file is plotted against original wave file matrix using plot () function.

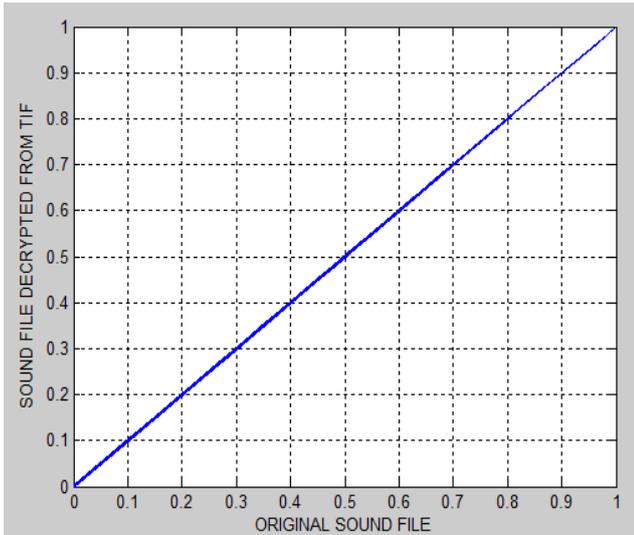

Figure 14 Plot of TIF decrypted sound file and original sound file.

The graph obtained is a straight line of type y=mx such that m=1. This implies that the sound file is exactly similar to the one obtained from TIF file format [15].
Using Equation A, we find that the error in this case is within a range of .0002 to -.0002.

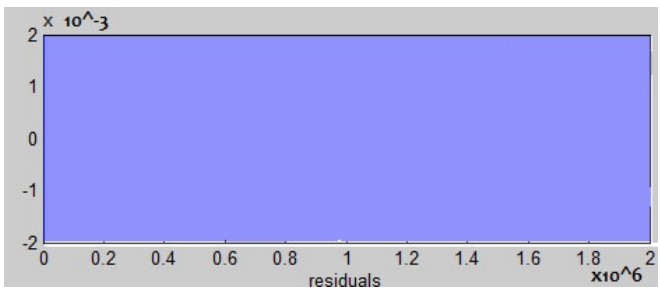

Figure 15 Range of error lies from -.0002 to .0002.

Thus TIF images encrypt sound files without compressing or hampering the data in them.

### 4.3 ORIGINAL WAVE FILE Vs. SOUND MATRIX OBTAINED FROM PNGFILE.

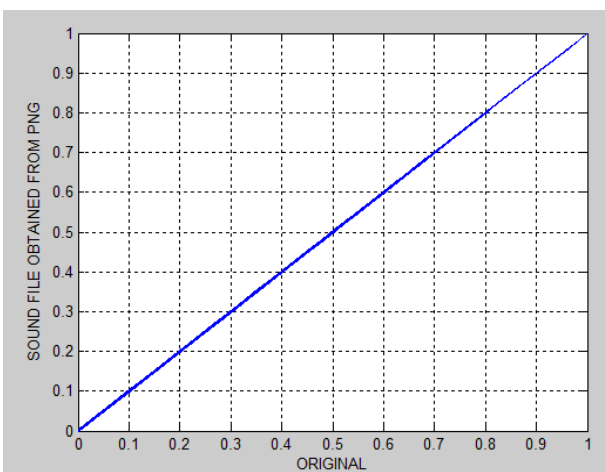

Figure 16 Plot of PNG decrypted sound file and original sound file.

Column matrix obtained from PNG file is plotted against original wave file matrix using plot () function. Error in this case is same as that in TIF decrypted sound.

### 5. DISCUSSION AND CONCLUSION

Following table summarises the results obtained in this paper.

TABLE IIII

|   | Image Type | Range of Error | Discussion |
|---|---|---|---|
| 1 | JPEG | -.05 to +.05 | JPEG file stores a sound file of 3.81 MB in less than 300kb but the sound quality is badly deteriorated. If sound quality can be compromised then JPEG can do best wave compression |
| 2 | TIF | -.0002 to .0002 | TIF format stores a 3.81 MB long wave file in around 1.6 MB with negligible error and the sound quality is perfect. |
| 3 | PNG | -.0002 to .0002 | PNG just like TIF stores data with negligible error however the space occupied by it is far less as compared to TIF. It stores sound files without hampering the sound quality to a significant extent. |

This method can be used for secured data transfer over networks wherein the entire image file containing the encrypted sound file would be arranged in a haphazard manner (or puzzled) and the retrieving end could be provided with the algorithm to extract and the sound file from the complex array [10].

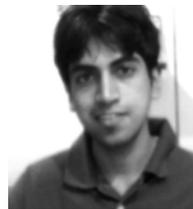

**Author Biography**

**Rahul R Upadhyay** is a B.Tech student of Department of Mechanical engineering, BBD National Institute of Technology and Management. He has an interdisciplinary research interest in topics varying from embedded electronics, network security, fluid mechanics and thermodynamics. He has published papers in National as well as International Journals and participated and won several technical awards in national level competitions and conferences.